\documentclass[graybox]{svmult}

\usepackage{mathptmx}       
\usepackage{helvet}         
\usepackage{courier}        
\usepackage{type1cm}        
\usepackage{lmodern}
\usepackage{natbib}
\usepackage{graphicx}        
\usepackage{multicol}        
\usepackage[bottom]{footmisc}

\usepackage{multirow}
\usepackage[table]{xcolor}
\usepackage{longtable}
\usepackage{tabu}
\usepackage{textpos}
\usepackage{amsmath}
\usepackage{amsfonts}
\usepackage{amssymb}
\usepackage{cases}
\usepackage{array}
\usepackage{color}
\usepackage{colortbl}
\usepackage{mdwlist}
\usepackage[it]{subfigure}
\usepackage{lscape}
\usepackage{lineno}
\usepackage{caption}

\makeindex             

\begin{document}

\title*{Introduction to Geodetic Time Series Analysis}
\titlerunning{Introduction to Geodetic Time Series Analysis}
\author{M.S. Bos, J.-P. Montillet, S.D.P. Williams, R.M.S. Fernandes}
 \authorrunning{Filtering GPS Time Series and Common Mode Error Analysis}
\institute{ M.S. Bos \at Instituto Dom Luiz, Universidade da Beira Interior, Portugal  \email{machiel@segal.ubi.pt}
\and J.-P. Montillet \at Space and Earth Geodetic Analysis Laboratory, Universidade da Beira Interior, Portugal -  Institute of Earth Surface Dynamics, University of Lausanne, Lausanne, Switzerland  \email{jpmontillet@segal.ubi.pt}
\and S.D.P. Williams \at National Oceanographic Centre, Liverpool, United Kingdom \email{sdwil@noc.ac.uk}
\and R.M.S. Fernandes \at Instituto Dom Luiz, Universidade da Beira Interior, Portugal  \email{rui@segal.ubi.pt}
}

\maketitle

\abstract{The previous chapter gave various examples of geophysical time series
and the various trajectory models that can be fitted to them. In this
chapter we will focus on how the parameters of the trajectory model can
be estimated. It is meant to give researchers new to this topic an easy
introduction to the theory with references to key books and articles where
more details can be found. In addition, we hope that it refreshes
some of the details for the more experienced readers. We pay special
attention to the modelling of the noise which has received much
attention in the literature in the last years and highlight some of the
numerical aspects. The subsequent chapters will go deeper into the
theory, explore different aspects and describe the state of art of this 
area of research.}

\section{Gaussian noise and the likelihood function}
\label{gaussian-noise-and-the-likelihood-function}

Geodetic time series consist out of a set observations at various
epochs. These observations, stored in a vector \(\vec y\), are not
perfect but contain noise which can be described as a set of
multivariate random variables. Let us define this as the vector
\({\vec w}=[W_1, W_2, W_3, \ldots, W_N]\) where each \(W_i\) is a random
variable. If \(f(w)\) is the associated probability density function,
then the first moment \(\mu_1\), the mean of the noise, is defined as
\citep{CasellaBerger2001}:
\begin{equation}
\label{firstmoment}
	\mu_1 = E[W] = \int\limits_{-\infty}^{\infty} w f(w)\:dw
\end{equation}

where \(E\) is the expectation operator. It assigns to each possible
value of random variable \(w\) a weight \(f(w)\) over an infinitely
small interval of \(dw\), sums each of them to obtain the mean expected
value \(E[W]\). The second moment \(\mu_2\) is defined in a similar
manner: 
\begin{equation}
\label{secondmoment}
	\mu_2 = E[W^2] = \int\limits_{-\infty}^{\infty} w^2 f(w)\:dw 
	= \int\limits_{-\infty}^{\infty} w^2 d F(w)
\end{equation}

The last term \(F\) is the cumulative distribution. The second moment is better
known as the variance. Since we have \(N\) random variables, we can
compute variances for \(E[W_i W_j]\), where both \(i\) and \(j\) range
from 1 to \(N\). The result is called the covariance matrix. 
In this book, the probability density function \(f(w)\) is assumed to be
a Gaussian:
\begin{equation}
	f(w|\mu_1,\sigma) = \frac{1}{\sqrt{2\pi\sigma^2}} 
	\text{exp}\left[-\frac{(w-\mu_1)^2}{2\sigma^2}\right]
\end{equation}

where \(\sigma\) is the standard deviation, the square-root of the
variance of random variable \(w\). This function is very well known and
is shown in Figure 1 for zero \(\mu_1\).

\begin{figure}
\centering
\includegraphics[scale=0.5]{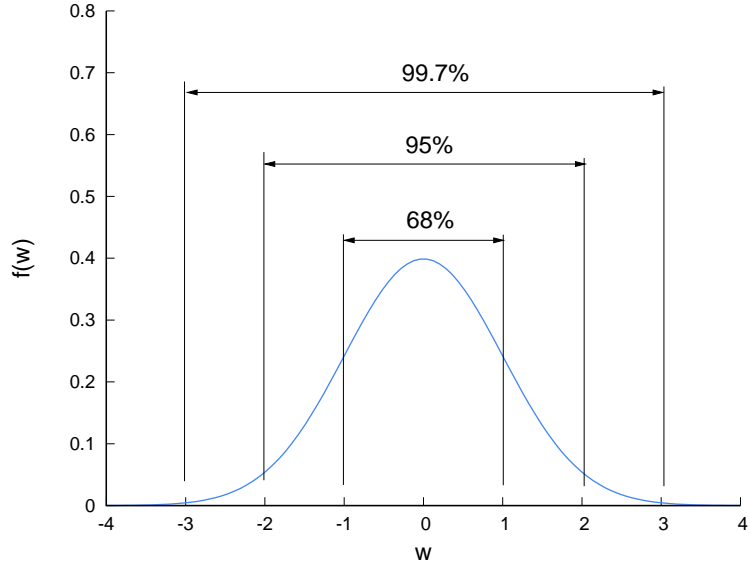}
\caption{The Gaussian probability density function, together with the 1,
2 and 3 \(\sigma\) intervals.}
\end{figure}

The standard error is defined as the 1-\(\sigma\) interval and contains
on average 68\% of the observed values of \(w\). The reason why it is so
often encountered in observations is that the central limit theorem
states that the sum of various continuous probability distributions
always tends to the Gaussian one. An additional property of the Gaussian
probability density function is that all its moments higher than two
(\(\mu_3\), \(\mu_4,\ldots\)) are zero. Therefore, the mean and the
covariance matrix provide a complete description of the stochastic
properties. Actually, we will always assume that the mean of the noise
is zero and therefore only need the covariance matrix. The term in front
of the exponential is needed to ensure that the integral of \(f(x)\)
from \(-\infty\) to \(\infty\) produces 1. That is, the total
probability of observing a value between these limits is 1, as it should
be. We have not one, but several observations with noise in our time
series. The probability density function of the multi-variate noise is:
\begin{equation}
	f({\vec w}|{\vec C}) = \frac{1}{\sqrt{(2\pi)^N\det({\vec C})}}
           \text{exp}\left[-\tfrac{1}{2}{\vec w}^T{\vec C}^{-1}{\vec w}\right]
\end{equation}

We assumed that the covariance matrix \(\vec C\) is known. The
expression \(f({\vec w}|{\vec C})\) should be read as the probability
density function \(f\) for variable \(w\), for given and fixed
covariance matrix \(\vec C\). 
Next, we assume that our observations can
be described by our model \({\vec g}({\vec x},t)\), where \({\vec x}\) are
the parameters of the model and \(t\) the time. The observations are the
sum of our model plus the noise:
\begin{equation}
	{\vec y} = {\vec g}({\vec x},t) + {\vec w}\;\;\;\text{or}\;\;\; 
	{\vec w}={\vec y} - {\vec g}({\vec x},t)
\end{equation}

The noise \(\vec w\) is described by our Gaussian probability
density function with zero mean and covariance matrix \(\vec C\). 
The probability that we obtained
the actual values of our observations is:
\begin{equation}
	f({\vec y}|{\vec x},{\vec C}) = \frac{1}{\sqrt{(2\pi)^N \det({\vec C})}}
             \text{exp}\left[-\tfrac{1}{2}({\vec y} - {\vec g}({\vec x},t))^T{\vec C}^{-1}({\vec y} - {\vec g}({\vec x},t))\right]
\end{equation}

However, we don't know the true values of \(\vec x\) or the covariance
matrix \(\vec C\). We only know the observations. Consequently, we need
to rephrase our problem as follows: what values of \(\vec x\) and
\(\vec C\) would produce the largest probability that we observe
\(\vec y\)? Thus, we are maximising \(f({\vec x},{\vec C}|{\vec y})\) which
we call the likelihood function \(L\). Furthermore, we normally work
with the logarithm of it which is called the log-likelihood:
\begin{equation}
	\ln(L)= -\frac{1}{2}\left[N\ln(2\pi) + \ln(\det({\vec C})) + ({\vec y} - {\vec g}({\vec x},t))^T{\vec C}^{-1}({\vec y} - {\vec g}({\vec x},t))\right]
\end{equation}

We need to find value of $\vec x$ to maximise this function and the method
is therefore called Maximum Likelihood Estimation (MLE).
The change from \(f({\vec y}|{\vec x},{\vec C})\) to
\(f({\vec x},{\vec C}|{\vec y})\) is subtle. Assume that the covariance
matrix \(\vec C\) also depends on parameters that we store in vector
\(\vec x\). In this way, we can simplify the expression
\(f({\vec y}|{\vec x},{\vec C})\) to \(f({\vec y}|{\vec x})\). Bayes'
Theorem, expressed in terms of probability distributions gives us:
\begin{equation}
	f({\vec x}|{\vec y}) = \frac{f({\vec y}|{\vec x}) f({\vec x})}{f({\vec y})}
\end{equation}
  
where \(f({\vec y})\) and \(f({\vec x})\) are our prior probability
density function for the observations \(\vec y\) and parameters
\(\vec x\), respectively. These represent our knowledge about what
observations and parameter values we expect before the measurements were
made. If we don't prefer any particular values, these prior probability
density functions can be constants and they will have no influence on
the maximising of the likelihood function \(f({\vec x}|{\vec y})=L\).

Another subtlety is that we changed from random noise and fixed
parameter values of the trajectory model \(f({\vec y}|{\vec x})\) to fixed
noise and random parameters of the trajectory model
\(f({\vec x}|{\vec y})\). If the trajectory model is for example a linear
tectonic motion then this is a deterministic, fixed velocity, not a
random one. However, one should interpret \(f({\vec x}|{\vec y})\) as our
degree of trust, our confidence that the estimated parameters \(\vec x\)
are correct. See also \cite{Koch1990,Koch2007} and \cite{Jaynes2003}. The 
last one is particularly recommended to learn more about Bayesian statistics.

\section{Linear models}\label{linear-models}

So far we simply defined our trajectory model as \({\vec g}({\vec x},t)\).
An important class of models that are fitted to the observations are linear
models. These are defined as:
\begin{equation}
	{\vec g}({\vec x},t) = x_1g_1(t) + x_2g_2(t) + \ldots + x_Mg_M(t)
\end{equation}

where \(x_1\) to \(x_M\) are assumed to be constants. We can rewrite
this in matrix form as follows: 
\begin{equation}
  {\vec g}({\vec x},t) = 
    \begin{pmatrix} g_1(t_1) & g_2(t_1) & \dots & g_M(t_1) \\
                    g_1(t_2) & g_2(t_2) &            & g_M(t_2) \\
                \vdots &                                & \vdots \\
                   g_1(t_N) & g_2(t_N) &        & g_M(t_N) 
    \end{pmatrix}
  \begin{pmatrix} x_1  \\ \vdots \\ x_M \end{pmatrix} =
        {\vec A}{\vec x}
\end{equation}

Matrix \(\vec A\) is called the design matrix. From Chapter 1 we
know that tectonic motion or sea level rise can be modelled by a linear
trend (i.e.~the Standar Linear Trajectory Model). Thus \(g_1(t)\) is a
constant and \(g_2(t)\) a linear trend. This can be extended to a higher
degree polynomial to model acceleration for example. Next, in many cases
an annual and semi-annual signal is included as well. A periodic signal
can be described by its amplitude \(b_k\) and its phase-lag \(\psi_k\) with
respect to some reference epoch:
\begin{equation}
\begin{split}
  g(t) &= b_k\cos(\omega_k t - \psi_k) \\
       &= b_k\cos\psi \cos(\omega_k t) + b_k\sin\psi_k \sin(\omega_k t)\\
       &=c_k\cos(\omega_k t) + s_k\sin(\omega_k t)
\end{split}
\end{equation}

Since the unknown phase-lag \(\psi_k\) makes the function non-linear, one
must almost always estimate the amplitudes \(c_k\) and \(s_k\), see Chapter 1.
These parameters
are linear with functions \(\cos\) and \(\sin\), and derive from these
values the amplitude \(b_k\) and phase-lag \(\psi_k\).

Other models that can be included in \(g(t)\) are offsets and
post-seismic relaxation functions, see Chapter 1. An example of a
combination of all these models into a single trajectory model is shown
in Figure 2.

\begin{figure}
\centering
\includegraphics[scale=0.5]{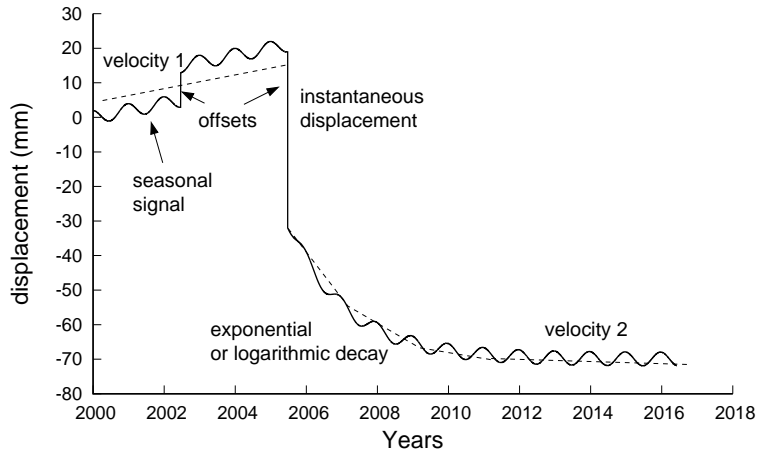}
\caption{Sketch of a trajectory model containing common phenomena}
\end{figure}

For linear models, the log-likelihood can be rewritten as:
\begin{equation}
	\ln(L)= -\frac{1}{2}\left[N\ln(2\pi) + \ln(\det({\vec C})) + ({\vec y} - {\vec A}{\vec x})^T{\vec C}^{-1}({\vec y} - {\vec A}{\vec x})\right]
\end{equation}

This function must be maximised. Assuming that the covariance matrix is
known, then it is a constant and does not influence finding the maximum.
Next, the term \(({\vec y} - {\vec A}{\vec x})\) represent the observations
minus the fitted model and are normally called the residuals \(\vec r\).
It is desirable to choosing the parameters \(\vec x\) in such a way to
make these residuals small. The last term can be written as
\({\vec r}^T{\vec C}^{-1}{\vec r}\) and it is a quadratic function,
weighted by the inverse of matrix \(\vec C\).

Now let us compute the derivative of \(\ln(L)\):
\begin{equation}
\label{deriv_lnL}
	\frac{d\ln(L)}{d{\vec x}} = {\vec A}^T{\vec C}^{-1} {\vec y} -
						{\vec A}^T{\vec C}^{-1}{\vec A}{\vec x} 	  
\end{equation}

The minimum of \(\ln(L)\) occurs when this derivative is zero. Thus:
\begin{equation}
\label{WLS}
	 {\vec A}^T{\vec C}^{-1}{\vec A}{\vec x} = {\vec A}^T{\vec C}^{-1} {\vec y} 
\;\;\rightarrow\;\; {\vec x}= \left({\vec A}^T{\vec C}^{-1}{\vec A}\right)^{-1}{\vec A}^T{\vec C}^{-1} {\vec y}
\end{equation}

This is the celebrated weighted least-squares equation to estimate the
parameters \(\vec x\). Most derivations of this equation focus on the
minimisation of the quadratic cost function. However, here we highlight
the fact that for observations that contain Gaussian multivariate noise,
the weighted least-squares estimator is a maximum likelihood estimator (MLE). 
From Eq. (\ref{WLS}) it can also be deduced that vector $\vec x$, like the
observation vector $\vec y$, follows a multi-variate Gaussian probability
density function.  

The variance of the estimated parameters estimated is:
\begin{equation}
\label{Cx}
\begin{split}
	var(\vec x) &= var\left(\left({\vec A}^T{\vec C}^{-1}{\vec A}\right)^{-1}
	{\vec A}^T{\vec C}^{-1}{\vec y}\right)\\
	  &= \left({\vec A}^T{\vec C}^{-1}{\vec A}\right)^{-1} {\vec A}^T{\vec C}^{-1} var({\vec y}) \; {\vec C}^{-1}  {\vec A} \left({\vec A}^T{\vec C}^{-1}{\vec A}\right)^{-1}\\
	  &= \left({\vec A}^T{\vec C}^{-1}{\vec A}\right)^{-1} {\vec A}^T{\vec C}^{-1} {\vec C} \; {\vec C}^{-1}  {\vec A} \left({\vec A}^T{\vec C}^{-1}{\vec A}\right)^{-1}\\
	  &= \left({\vec A}^T{\vec C}^{-1}{\vec A}\right)^{-1}
\end{split}
\end{equation}

Next, define the following matrix ${\mathcal I}({\vec x})$:
\begin{equation}
	{\mathcal I}({\vec x}) = -E\left[\frac{\partial^2}{\partial {\vec x}^2}
	   \ln(L)\right] = -\int\limits \left(\frac{\partial^2}{\partial 
	   {\vec x}^2} \ln(f)\right) \; f \;d{\vec x}
\end{equation}

It is called the Fisher Information matrix. As in Eqs. (\ref{firstmoment})
and (\ref{secondmoment}), we use the expectation operator $E$. Remember
that we simply called $f$ our likelihood $L$ but these are the same. 
We already used the fact
that the log-likelihood as function of $\vec x$ is horizontal at the
maximum value. Let us call this $\hat{\vec x}$. 
The second derivative is related to the curvature of the
log-likelihood function. The sharper the peak near its maximum, 
the more accurate we can estimate the parameters $\vec x$ and therefore 
the smaller their variance will be.

Next, it can be shown that the following inequality holds:
\begin{equation}
 1 \le 	\int\limits (\hat{\vec x} - {\vec x})^2\; f \;d{\vec x}
  \int\limits \left(\frac{\partial \ln(f)}
  {\partial {\vec x}}\right)^2\; f\;d{\vec x}
\end{equation}

The first integral represents the variance of $\vec x$, see Eq. 
(\ref{secondmoment}). The second one, after some rewriting, is equal to 
the Fisher information matrix. This gives us, for any unbiased estimator, 
the following Cram\'er-Rao Lower Bound \citep{Kay1993}:
\begin{equation}
\label{CRLB}
	var(\hat{\vec x}) \geq \frac{1}{{\mathcal I}({\vec x})}
\end{equation}

Eq. ({\ref{CRLB}) predicts the minimum variance of the estimated
parameters $\vec x$ for given probability density function $f$ and its
relation with the parameters $\vec x$ that we want to estimate. If we
us Eq. (\ref{deriv_lnL}) to compute the second derivative of the
log-likelihood, then we obtain:
\begin{equation}
	{\mathcal I}({\vec x}) = {\vec A}^T{\vec C}^{-1}{\vec A}
\end{equation} 

Comparing this with Eq. (\ref{Cx}), one can see that for the case of
the weighted least-square estimator, the Cram\'er-Rao Lower Bound is 
achieved. Therefore, it is an optimal estimator. Because we also need to 
estimate the parameters of the covariance matrix $\vec C$, we shall use
MLE which approximates this lower bound for increasing
number of observations. Therefore, one can be sure that out of all existing
estimation methods,
none of them will produce a more accurate result than MLE, only equal or 
worse. For more details, see \cite{Kay1993}.

\section{Models for the covariance matrix}
\label{models-for-the-covariance-matrix}

Least-squares and maximum likelihood estimation are well known techniques in
various branches of science. In recent years much attention has been paid
by geodesists to the structure of the covariance matrix. If there was no
relation between each noise value, then these would be independent
random variables and the covariance matrix \(\vec C\) would be zero
except for values on its diagonal. However, in almost all geodetical
time series these are dependent random variables. In statistics this is
called temporal correlation and we should consider a full covariance
matrix: 
\begin{equation}
\label{samplecovariance}
	{\vec C} = \begin{pmatrix} 
	\sigma^2_{11} & \sigma^2_{12} & \ldots & \sigma^2_{1N} \\
    \sigma^2_{21} & \sigma^2_{22} &     & \sigma^2_{2N} \\
     \vdots & & \ddots & \vdots \\
     \sigma^2_{N1} & \ldots & \sigma^2_{N N-1} & \sigma^2_{NN} 
  \end{pmatrix}
\end{equation}
  
where \(\sigma^2_{12}\) is the covariance between random variables
\(w_1\) and \(w_2\). If we assume that the properties of the noise are
constant over time, then we have the same covariance between \(w_2\) and 
\(w_3\), \(w_3\) and \(w_4\) and all other correlations with 1 time step
separation. As a result, \(\sigma^2_{12}\), \(\sigma^2_{23},\ldots\), 
\(\sigma^2_{(N-1)N}\) are all equal. A simple estimator for it is:
\begin{equation}
	\sigma^2_{12}= \sigma^2_{23} = \ldots = \sigma^2_{(N-1)N} = \frac{1}{N-1}\sum\limits_{i=1}^{N-1} w_i w_{i+1}
\end{equation}

This is an approximation of the formula to compute the second moment,
see Eq. (\ref{secondmoment}), and it called the empirical or sample 
covariance matrix.
Therefore, one could try the following iterate scheme: fit the linear
model to the observations some a priori covariance matrix, compute the
residuals and use this to estimate a more realistic covariance matrix
using Eq. (\ref{samplecovariance}) and fit again the linear model to the
observations until all estimated parameters have converged.

The previous chapter demonstrated that one of the purpose of the trajectory
models is to estimate the linear or secular trend. For time series longer than 2 years, the uncertainty of
this trend depends mainly on the noise at the lowest observed periods 
\citep{Bosetal2008,Heetal2019}}. However, the empirical covariance matrix
 estimation of Eq. (\ref{samplecovariance})
does not result in an accurate estimate of the noise at  long periods
because only a few observations are used in the computation. In fact,
only the first and last observation are used to compute the variance of the 
noise at the longest observed period (i.e. \(\sigma^2_{1N}\)).

This problem has been solved by defining a model of the noise and
estimating the parameters of this noise model. The estimation of the
noise model parameters can be achieved using the log-likelihood with a
numerical maximisation scheme but other methods exist such as
least-squares variance component estimation (see Chapter 6).

The development of a good noise model started with the paper of \cite{Hurst1957}
who discovered that the cumulative water flow of the Nile river
depended on the previous years. The influence of the previous years
decayed according a power-law. This inspired \cite{MandelbrotNess1968}
to define the fractional Brownian motion model which includes both the 
power-law and fractional Gaussian noises, see also \cite{Beran1994} and
 \cite{Gravesetal2017}.
While this research was well known in hydrology and in
econometry, it was not until the publication by \cite{Agnew1992}, who
demonstrated that most geophysical time series exhibit power-law noise
behaviour, that this type of noise modelling started to be applied to
geodetic time series. In hindsight, \cite{Press1978} had already
demonstrated similar results but this work has not received much
attention in geodesy. That the noise in GNSS time series also falls in
this category was demonstrated by \cite{JohnsonAgnew1995}. Power-law
noise has the property that the power spectral density of the noise
follows a power-law curve. On a log-log plot, it converts into a
straight line. The equation for power-law noise is:
\begin{equation}
\label{powerlaw}
  P(f) = P_0\:(f/f_s)^\kappa
\end{equation}

where \(f\) is the frequency, \(P_0\) is a constant, \(f_s\) the
sampling frequency and the exponent \(\kappa\) is called the spectral
index.

\cite{Granger1980}, \cite{GrangerJoyeux1980} and \cite{Hosking1981}
demonstrated that power-law noise can be achieved using fractional
differencing of Gaussian noise: 
\begin{equation}
\label{fractionaldif}
	(1-B)^{-\kappa/2} {\vec v} = {\vec w}
\end{equation}

where \(B\) is the backward-shift operator (\(B v_i = v_{i-1}\)) and
\(\vec v\) a vector with independent and identically distributed (IID)
Gaussian noise. Hosking and Granger used the parameter \(d\) for the
fraction \(-\kappa/2\) which is more concise when one focusses on the
fractional differencing aspect. It has been adopted by people studying
general statistics \citep{Sowell1992,Beran1995}. However, in geodesy the
spectral index \(\kappa\) is used in the equations. Hosking's definition
of the fractional differencing is:
\begin{equation}
\label{Hoskingdef}
\begin{split}
 (1-B)^{-\kappa/2} &= \sum\limits_{i=0}^\infty \binom{-\kappa/2}{i}(-B)^i \\
 &= 1- \frac{\kappa}{2}B  -  \frac{1}{2}\frac{\kappa}{2}(1-\frac{\kappa}{2})B^2+\ldots\\
 &= \sum\limits_{i=0}^\infty h_i
\end{split}
\end{equation}

The coefficients \(h_i\) can be viewed as a filter that is applied to
the independent white noise. These coefficients can be conveniently
computed using the following recurrence relation \citep{Kasdin1995}:
\begin{equation}
\label{h_recurrence}
\begin{split}
h_0 &= 1 \\
h_i &= (i-\frac{\kappa}{2}-1)\frac{h_{i-1}}{i}\;\;\;\;\text{ for }i>0
\end{split}
\end{equation}

One can see that for increasing \(i\), the fraction \((i-\kappa/2-1)/i\)
is slightly less than 1. Thus, the coefficients \(h_i\) only decrease
very slowly to zero. This implies that the current noise value \(w_i\)
depends on many previous values of \(\vec v\). In other words, the noise
has a long memory. 
Actually, the model of fractional Gaussian noise defined by 
\cite{Hosking1981} is the basic definition of the general class of processes
called Auto Regressive Integrated moving Average \citep{Taqquetal1995}. 
If we ignore the Integrated part, then we obtain the Auto
Regressive Moving Average (ARMA) model 
\citep{Boxetal2015,BrockwellDavis2002} which are short-memory noise models.
The original definition of the ARIMA processes only considers the value of 
the power $\kappa/2$ in Eq. (\ref{Hoskingdef}) as an integer value. 
\cite{GrangerJoyeux1980} further extended the definition to a class of
fractionally integrated models called FARIMA or ARFIMA, where $\kappa$ is a 
floating value, generally in the range of $-1 \le  i \le 1$. 
\cite{MontilletYu2015} discussed the application of the ARMA and FARIMA 
models in modelling GNSS daily position time series and concluded that the
FARIMA is only suitable in the presence of a large amplitude coloured noise
capable of generating a distribution with large tails (i.e. random-walk,
aggregations).

Equation (\ref{h_recurrence}) also shows that when the spectral index 
\(\kappa=0\), then
all coefficients \(h_i\) are zero except for \(h_0\). This implies that
there is no temporal correlation between the noise values. In addition,
Eq. (\ref{powerlaw}) shows that this corresponds to a horizontal line in the 
power spectral density domain. Using the analogy of the visible light
spectrum, this situation of equal power at all frequencies produces
white light and it is therefore called white noise. For \(\kappa\ne0\),
some values have received a specific colour. For example, \(\kappa=-1\)
is known as pink noise. Another name is flicker noise which seems to
have originated in the study of noise of electronic devices. Red noise
is defined as power-law noise with \(\kappa=-2\) and produces \(h_i=1\)
for all values of \(i\). Thus, this noise is a simple sum of all its
previous values plus a new random step and is better known as random
walk \citep{Mandelbrot1999}. However, note that the spectral index $\kappa$
does not need to be an integer value \citep{Williams2003}.

One normally assumes that \(v_i=0\) for \(i<0\). With this assumption,
the unit covariance between \(w_k\) and \(w_l\) with \(l>k\) is:
\begin{equation}
\label{powerlawcovsum}
	C(w_k,w_l) = \sum\limits_{i=0}^k h_i h_{i+(l-k)}
\end{equation}

Since \(\kappa=0\) produces an identity matrix, the associated white
noise covariance matrix is represented by unit matrix \(\vec I\). The general
power-law covariance matrix is represented by the matrix \(\vec J\). The sum of
white and power-law noise can be written as \citep{Williams2003}:
\begin{equation}
	{\vec C} = \sigma_{pl}^2 {\vec J}(\kappa) + \sigma_w^2 {\vec I}
\end{equation}
  
where \(\sigma_{pl}\) and \(\sigma_w\) are the noise amplitudes. It
is a widely used combination of noise models to describe the noise in
GNSS time series (Williams et al. 2004). Besides the parameters of the
linear model (i.e.~the trajectory model), maximum likelihood estimation
can be used to also estimate the parameters \(\kappa\), \(\sigma_{pl}\)
and \(\sigma_w\). This approach has been implemented various software
packages such as CATS \citep{Williams2008}, est\_noise \citep{Langbein2010} and
Hector (Bos et al. 2013). In recent years one also has detected random
walk noise in the time series and this type has been included as well in
the covariance matrix \citep{Langbein2012,Dmitrievaetal2015}.

We assumed that \(v_i=0\) for \(i<0\) which corresponds to no noise
before the first observation. This is an important assumption that has
been introduced for a practical reason. For a spectral index \(\kappa\)
smaller than -1, the noise becomes non-stationary. That is, the variance
of the noise increases over time. If it is assumed that the noise was 
always present, then the variance would be infinite.

Most GNSS time series contain flicker noise which is just non-stationary. 
Using the
assumption of zero noise before the first observation, the covariance
matrix still increases over time but remains finite.

For some geodetic time series, such as tide gauge observations, the
power-law behaviour in the frequency domain shows a flattening below
some threshold frequency. To model such behaviour, \cite{Langbein2004}
introduced the Generalised Gauss-Markov (GGM) noise model which is
defined as:
\begin{equation}
	(1-\phi B)^{-\kappa/2} {\vec v} = {\vec w}
\end{equation}

The only new parameter is \(\phi\). The associated recurrence relation
to compute the new coefficients \(h_i\) is: 
\begin{equation}
\begin{split}
h_0 &= 1 \\
h_i &= (i-\frac{\kappa}{2}-1)\phi\frac{h_{i-1}}{i}\;\;\;\;\text{ for }i>0
\end{split}
\end{equation}

If \(\phi=1\), then we obtain again our pure power-law noise model. For
any value of \(\phi\) slightly smaller than one, this term helps to
shorten the memory of noise which makes it stationary. That is, the
temporal correlation decreases faster to zero for increasing lag between
the noise values. The power-spectrum of this noise model shows a
flattening below some threshold frequency which guarantees that the
variance is finite and that the noise is stationary. Finally, it is even
possible to generalise this a bit more to a fractionally integrated
generalised Gauss-Markov model (FIGGM): 
\begin{equation}
\begin{split}
(1-\phi B)^{-\kappa_1/2} (1-B)^{\kappa_2/2} {\vec v} &= {\vec w} \\
  (1-\phi B)^{-\kappa_1/2}  {\vec u} = {\vec w}
\end{split}
\end{equation}

This is just a combination of the two previous models. One can first
apply the power-law filter to \(\vec v\) to obtain \(\vec u\) and
afterwards apply the GGM filter on it to obtain \(\vec w\). Other models
will be discussed in this book, such as ARMA
\citep{Boxetal2015,BrockwellDavis2002}, but the power-law, 
GGM and FIGGM capture 
nicely the long memory property that is present in most geodetic time series. 
A list of all these noise models and their abbreviation is given in 
Table \ref{noisemodelnames}.

\begin{table}
\caption{Common abbreviation of noise models}
\label{noisemodelnames}
\begin{tabular}{ll}
\hline
Noise model & Abbreviation	\\ \hline
Auto-Regressive Moving Average & ARMA \\
Auto-Regressive Fractionally Integrated Moving Average & ARFIMA or FARIMA\\
Flicker noise                  & FN \\
Fractionally Integrated GGM    & FIGGM \\
Generalised Gauss Markov       & GGM \\
Power-law                      & PL \\
Random Walk                    & RW \\
White noise                    & WN \\
\hline
\end{tabular}	
\end{table}

\section{Power spectral density}
\label{power-spectral-density}

\begin{figure}
\includegraphics[scale=0.55]{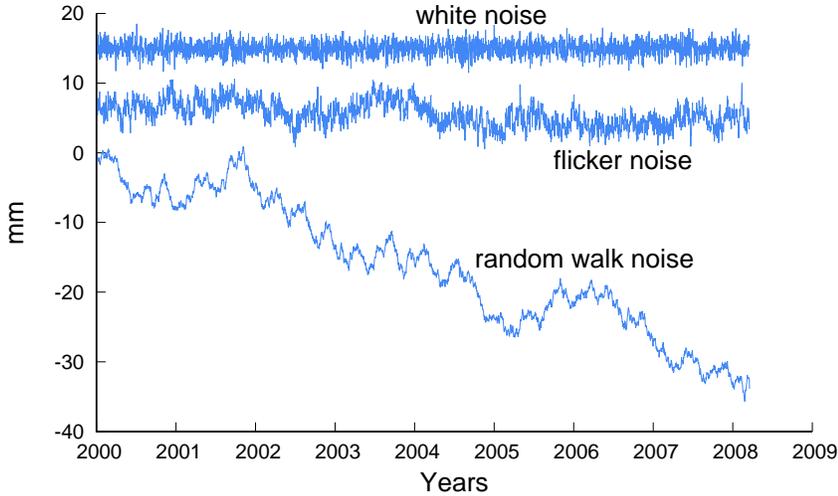}
\caption{\label{powerlawexamples}
Examples of white, flicker and random walk noise}
\end{figure}

Figure \ref{powerlawexamples} shows examples of white, flicker and 
random walk noise for a displacement time series. One can
see that the white noise varies around a stable mean
while the random walk is clearly non-stationary and deviates away from
its initial position.

In the previous section we mentioned that power-law noise has a specific
curve in the power spectral density plots. Methods to compute those 
plots are given by \cite{Buttkus2000}. A simple but effective method
is based on the Fourier transform that states that 
each time series with finite variance can be written as a sum of 
periodic signals:
\begin{equation}
\label{invfft}
	y_n = \frac{1}{N}\sum\limits_{k=-N/2+1}^{N/2} Y_k \cdot e^{i2\pi kn/N}\;\;\;\text{ for }n=[0,\dots,N-1]
\end{equation}

Actually, this is called the inverse discrete Fourier transform. \(Y_k\) are
complex numbers, denoting the amplitude and phase of the periodic signal with
period \(k/(NT)\) where \(T\) is the observation span. An attentive
reader will remember that flicker and random walk noise are non-stationary
while the Fourier transform requires time series with finite variance. 
However, we never have
infinitely long time series which guarantees the variance remains within 
some limit. 
The coefficients can be computed as follows:
\begin{equation}
\label{fft}
	Y_k = \sum\limits_{n=0}^{N-1} y_n \cdot e^{-i2\pi kn/N}\;\;\;
										\text{ for }k=[-N/2+1,\ldots,N/2]
\end{equation}

The transformation to the frequency domain provides insight which
periodic signals are present in the signal and in our case, insight
about the noise amplitude at the various frequencies. This is a classic
topic and more details can be found in the books by \cite{Bracewell1978} and
\cite{Buttkus2000}. The one-sided power spectral density \(S_k\) is defined as:
\begin{equation}
\begin{split}
	S_0 &= |Y_0|^2/f_s \\
	S_{N/2} &=  |Y_{N/2}|^2/f_s \\
	S_k &= 2|Y_k|^2/f_s \;\;\;\text{ for }k=[1,\ldots,N/2-1]
\end{split}
\end{equation}

The frequency $f_k$ associated to each $S_k$ is:
\begin{equation}
  f_k = \frac{ k f_s}{N}	\;\;\;\text{ for }k=[0,\ldots,N/2]
\end{equation}

The highest frequency is half the sampling frequency, $f_s/2$, which 
is called the Nyquist frequency. The power spectral density
(PSD) computed in this manner is called a periodogram. There are many
refinements, such as applying window functions and cutting the time
series in segments and averaging the resulting set of PSD's. However, a
detail that normally receives little attention is that the Fourier
transform produces positive and negative frequencies. Time only
increases and there are no negative frequencies. Therefore, one always
uses the one-sided power spectral density. Another useful
relation is that of Parseval \citep{Buttkus2000}:
\begin{equation}
	\frac{1}{N}\sum\limits_{n=0}^{N-1} |y_n|^2 = 
						\frac{1}{N^2}\sum\limits_{k=-N/2+1}^{N/2} |Y_k|^2
\end{equation}

\begin{figure}
\includegraphics[scale=0.4]{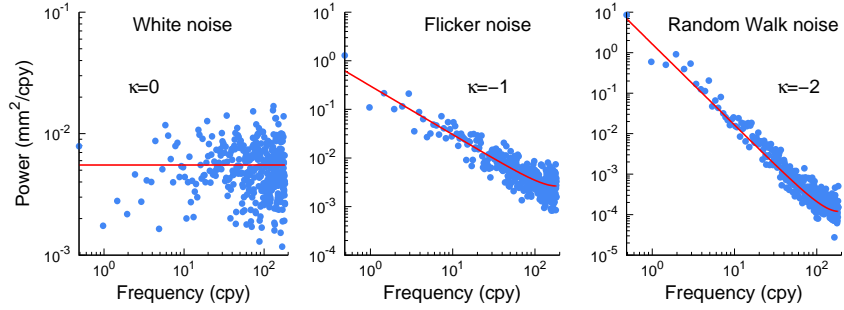}
\caption{\label{periodogram}
One-sided power spectral density for white, flicker and random
walk noise. The blue dots are the computed periodogram (Welch's method)
while the solid red line is the fitted power-law model.}
\end{figure}

Thus, the variance of the noise should be equal to the sum of all
\(S_k\) values (and an extra \(f_s/N^2\) scale). The one-sided power
spectral density of the three time series of Figure \ref{powerlawexamples}
are plotted in Figure \ref{periodogram}. It shows that power-law noise indeed
follows a straight line in the power spectral density plots if a log-log
scale is used. In fact, the properties of the 
power-law noise can also be
estimated by fitting a line to the power spectral density estimates
\citep{Maoetal1999,Caporali2003}.

The PSD of power-law noise generated by fractionally differenced
Gaussian noise is \citep{Kasdin1995}:
\begin{equation}
\begin{split}
	S(f) &= \frac{2\sigma^2}{f_s}(2\sin(\pi f/f_s))^{\kappa} \\
	&\approx \frac{2\sigma^2}{f_s} (\pi f/f_s))^{\kappa} =
					 P_0 (f/f_s)^\kappa \;\;\;\text{ for }f\ll f_s
\end{split}
\end{equation}

For small value of \(f\), this approximates
\(P_0 (f/f_s)^\kappa\). The sine function is the result of having discrete
data \citep{Kasdin1995}. The PSD for GGM noise is: 
\begin{equation}
   S(f) = \frac{2\sigma^2}{f_s}
			\left[1+\phi^2-2\phi\cos(2\pi f/f_s)\right]^{\kappa/2}
\end{equation}

For \(\phi=1\), it converts to the pure power-law noise PSD.
The Fourier transform, and especially the Fast Fourier Transform, can
also be used to filter a time series. For example, Eqs. (\ref{fractionaldif})
and (\ref{Hoskingdef}) represent a filtering of white noise vector $\vec v$
to produce coloured noise vector $\vec w$:
\begin{equation}
\label{colourednoise}
	w_i = \sum\limits_{j=0}^{i-1} h_{i-j} \; v_j
\end{equation}

Let us now extend the time series $\vec y$ and the vector $\vec h$ containing
the filter coefficients with $N$ zeros. This zero padding
allows us to extend the summation to $2N$. Using Eq. (\ref{fft}), their Fourier
transforms, $Y_k$ and $H_k$, can be computed. In the frequency domain,
convolution becomes multiplication and we have \citep{Pressetal2007}:
\begin{equation}
	W_k = H_k \;Y_k\;\;\;\text{ for }k=[-N,\ldots,N]
\end{equation}

Using Eq. ({\ref{invfft}) and only using the first $N$ elements, 
the vector $\vec w$ with the coloured noise can be obtained.

\section{Numerical examples}
\label{numericalexamples}

To explain the principle of maximum likelihood, this section will show some
examples of the numerical method using Python 3. For some years Matlab has been
the number one choice to analyse and visualise time series. However, in recent
years Python has grown in popularity, due to the fact that 
it is open source and has
many powerful libraries. The following examples are made in IPython 
(https://ipython.org), using the Jupyter notebook webapplication. How to 
install this program is described on the afore mentioned website. The examples
shown here can be downloaded from the publisher website.
The first step is to import the libraries:
\begin{verbatim}
import math
import numpy as np
from matplotlib import pyplot as plt
from scipy.optimize import minimize
from numpy.linalg import inv	
\end{verbatim}

Next step is to create some data which we will store in Numpy arrays. As in
Matlab, the `linspace' operator creates a simple array on integers. Furthermore,
as the name implies 'random.normal' creates an array of Gaussian distributed
random numbers. We create a line $\vec y$ with slope 2 and offset 6 on which we 
superimpose the noise $\vec w$ that were created using a standard deviation
$\sigma_{pl}=0.5$ for vector $\vec v$, see Eq. (\ref{fractionaldif}).
\begin{verbatim}
N = 500                        # Number of daily observations
t = np.linspace(0,N/365.25,N)  # time in years
np.random.seed(0)              # Assure we always get the same noise

kappa = -1               #  Flicker noise
h     = np.zeros(2*N)    #  Note the size : 2N
h[0]  = 1                #  Eq. (25)
for i in range(1,N):
    h[i] = (i-kappa/2-1)/i * h[i-1]
    
v      = np.zeros(2*N)   # Again zero-padded N:2N
v[0:N] = np.random.normal(loc = 0.0, scale = 0.5, size = N)

w = np.real(fft.ifft(fft.fft(v) * fft.fft(h)))  # Eq. (39)

y = (6 + 3*t) + w[0:N]   # trajectory model + noise

plt.plot(t, y, 'b-')     # plot the time series    
\end{verbatim}

\begin{figure}
\includegraphics[scale=0.8]{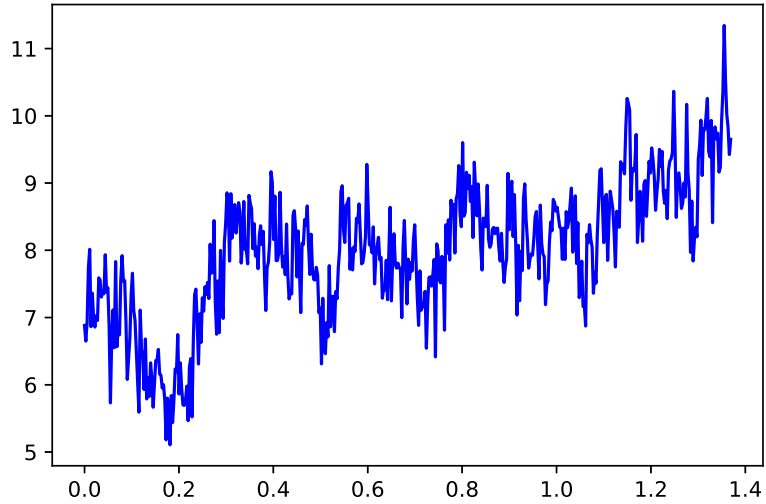}
\caption{\label{synthetic_ts}
Our synthetic time series containing a simple line plus flicker noise.}
\end{figure}	

Of course the normal situation is that we are given a set observations and
that we need to estimate the parameters of the trajectory model $y(t)=a+bt$.
However,
creating synthetic time series is a very good method to test if your
estimation procedures are correct.

First we will estimate the trajectory assuming white noise in the data:
\begin{verbatim}
#--- The design matrix
A = np.empty((N,2))
for i in range(0,N):
    A[i,0] = 1
    A[i,1] = t[i]
    
#--- Old white noise method
C     = np.identity(N)
x     = inv(A.T @ inv(C) @ A) @ (A.T @ inv(C) @ y)  # Eq. (14)
y_hat = A @ x
r     = y - y_hat                                   # residuals
C_x   = np.var(r)* inv(A.T @ inv(C) @ A)            # Eq. (15)
print('White noise approximation')
print('a = {0:6.3f} +/- {1:5.3f} mm'.format(x[0],\
                                          math.sqrt(C_x[0,0])))
print('b = {0:6.3f} +/- {1:5.3f} mm/yr'.format(x[1],\
                                          math.sqrt(C_x[1,1])))
\end{verbatim}}

The result should be:
\begin{verbatim}
White noise approximation
a =  6.728 +/- 0.064 mm
b =  1.829 +/- 0.080 mm/yr
\end{verbatim}

What we have done here is using weighted least-squares with a white noise
model that has unit variance. The real variance of the noise has been
estimated from the residuals and the uncertainty of the estimated
parameters $\vec x$ have been scaled with it.

At this point the reader will realise that this approach is not justified
because the noise is temporally correlated. It will be convenient to
define the following two functions that will create the covariance matrix
for power-law noise and apply weighted least-squares
\citep{Williams2003,Bosetal2008}:
\begin{verbatim}
#--- power-law noise covariance matrix
def create_C(sigma_pl,kappa):
    U      = np.identity(N)
    h_prev = 1              
    for i in range(1,N):
        h = (i-kappa/2-1)/i * h_prev #  Eq. (25)
        for j in range(0,N-i):
            U[j,j+i] = h
        h_prev = h
    U *= sigma_pl                    # scale noise
    return U.T @ U                   # Eq. (26)


#--- weighted least-squares 
def leastsquares(C,A,y):
    U     = np.linalg.cholesky(C).T
    U_inv = inv(U)
    B     = U_inv.T @ A
    z     = U_inv.T @ y
    x     = inv(B.T @ B) @ B.T @ z           # Eq. (14)

    #--- variance of the estimated parameters
    C_x = inv(B.T @ B)                       # Eq. (15)
    
    #--- Compute log of determinant of C
    ln_det_C = 0.0
    for i in range(0,N):
        ln_det_C += 2*math.log(U[i,i])
    
    return [x,C_x,ln_det_C]
\end{verbatim}

The function that creates the covariance matrix for power-law noise
has been discussed in section \ref{models-for-the-covariance-matrix} and 
uses Eqs. (\ref{h_recurrence}) and (\ref{powerlawcovsum}). The
weighted least-squares function
contains some numerical tricks. First, the Cholesky decomposition is applied
to the covariance matrix \citep{Bosetal2008}:
\begin{equation}
	{\vec C} = {\vec U}^T{\vec U} 
\end{equation}

where $\vec U$ is an upper triangle matrix. That is, only the elements above
the diagonal are non-zero. A covariance matrix is a positive definite matrix 
which ensures that the Cholesky decomposition always exists. The most important
advantage it that one can compute the logarithm of the determinant of 
matrix $\vec C$ by just summing the logarithm of each element on the diagonal
of matrix $\vec U$. The factor two is needed because matrix $\vec C$ is the
product of ${\vec U}^T{\vec U} $. Using these two functions, we can compute
the correct parameters $\vec x$:
\begin{verbatim}
#--- The correct flicker noise covariance matrix
sigma_pl         = 4
kappa            = -1
C                = create_C(sigma_pl,kappa)
[x,C_x,ln_det_C] = leastsquares(C,A,y)
print('Correct Flicker noise')
print('a = {0:6.3f} +/- {1:5.3f} mm'.format(x[0],\
                                        math.sqrt(C_x[0,0])))
print('b = {0:6.3f} +/- {1:5.3f} mm/yr'.format(x[1],\
                                        math.sqrt(C_x[1,1])))	
\end{verbatim}

The result is:
\begin{verbatim}
Correct Flicker noise
a =  6.854 +/- 2.575 mm
b =  1.865 +/- 4.112 mm/yr
\end{verbatim}

If one compares the two estimates, one assuming white noise and the other
assuming flicker noise, then one can verify that the estimates themselves
are similar. The largest difference occurs for the estimated errors which
are 5 times larger for the latter. This also happens in real geodetic time
series. \cite{Maoetal1999} concluded that the velocity error in GNSS 
time-series could be underestimated by factors of 5–11 if a pure white 
noise model is assumed. \cite{Langbein2012} demonstrated that an additional
factor of two might be needed if there is also random walk noise present.

For sea level time series \cite{Bosetal2014} obtained a more moderate 
factor of 1.5-2 but still, white noise underestimates the true uncertainty
of the estimated linear trend. \cite{Williamsetal2014} estimated a factor 6
for the GRACE gravity time series. As discussed in section 
\ref{models-for-the-covariance-matrix}, most geodetic time series are 
temporally correlated and therefore one nowadays avoids the white noise model.

So far we have assumed that we knew the true value of the spectral 
index $\kappa$ and the noise amplitude $\sigma_{pl}$. Using MLE, we 
can estimate these parameters from the data:
\begin{verbatim}
#--- Log-likelihood (with opposite sign)
def log_likelihood(x_noise):
    sigma_pl = x_noise[0]
    kappa    = x_noise[1]
    C        = create_C(sigma_pl,kappa)
    [x,C_x,ln_det_C] = leastsquares(C,A,y)
    r        = y - A @ x  # residuals
    
    #--- Eq. (12)
    logL     = -0.5*(N*math.log(2*math.pi) + ln_det_C \
                                        + r.T @ inv(C) @ r)
    return -logL


x_noise0 = np.array([1,1]) # sigma_pl and kappa guesses
res = minimize(log_likelihood, x_noise0, \
               method='nelder-mead', options={'xatol':0.01})

print('sigma_pl={0:6.3f}, kappa={1:6.3f}'.\
                                  format(res.x[0],res.x[1]))
\end{verbatim}

Note that we inverted the sign of the log-likelihood function because
most software libraries provide minisation subroutines, not maximisation. 
In addition,
it is in this function that we need the logarithm of the determinant of
matrix $\vec C$. If one tries to compute it directly from matrix $\vec C$,
then one quickly encounters too large numbers that create numerical overflow.
This function also shows that we use weighted least-squares to estimate
the parameters of the trajectory model while the numerical minisation 
algorithm (i.e. Nelder-Mead), is only used the compute the noise parameters.
The reason for using weighted least-squares, also a maximum likelihood
estimator as we have shown in section \ref{linear-models}, 
is solely for speed. Numerical
minisation is a slow process which becomes worse for each additional
parameter we need to estimate.
The results is:
\begin{verbatim}
sigma_pl= 0.495, kappa=-1.004
\end{verbatim}

These values are close to the true values of $\sigma_{pl}=0.5$ and 
$\kappa=-1$. The following code can be used to plot the log-likelihood
as function of $\kappa$ and $\sigma_{pl}$:
\begin{verbatim}
S = np.empty((21,21))
for i in range(0,21):
    sigma_pl = 1.2 - 0.05*i
    for j in range(0,21):
        kappa = -1.9 + 0.1*j
        x_noise0 = [sigma_pl,kappa]
        S[i,j] = math.log(log_likelihood(x_noise0))
        
plt.imshow(S,extent=[-1.9,0.1,0.2,1.2], cmap='nipy_spectral', \
                                               aspect='auto');
plt.colorbar()
plt.ylabel('sigma_pl')
plt.xlabel('kappa')
plt.show()
\end{verbatim}

\begin{figure}
\includegraphics[scale=0.9]{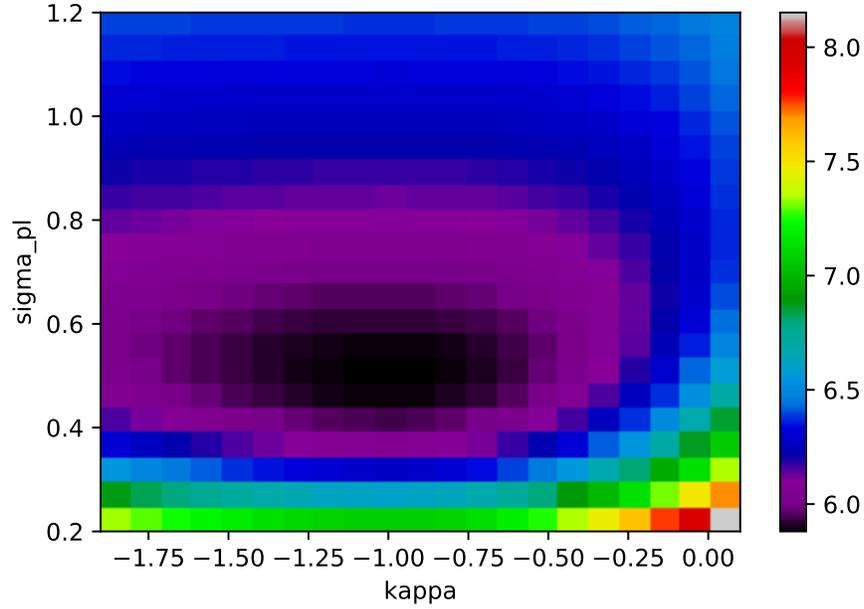}
\caption{\label{minimum} The log of the $\log(L)$ function as function of
$\kappa$ and $\sigma_{pl}$.}
\end{figure}

The result is shown in Figure \ref{minimum} which indeed shows a minimum
around $\sigma_{pl}=0.5$ and $\kappa=-1$. Depending on the computer power,
it might take some time to produce the values for this figure. 

In section \ref{models-for-the-covariance-matrix} we noted that for GNSS
time series the power-law plus white noise model is common. Thus, one
must add the covariance matrix for white noise, $\sigma_w^2{\vec I}$, to
the covariance matrix we discussed in the examples. In addition,
it is more efficient to write the covariance matrix of
the sum of power-law and white noise as follows:
\begin{equation}
\label{SimonRewrite}
	{\vec C} = \sigma_{pl}^2{\vec J}(\kappa) + \sigma_w^2{\vec I}=
			\sigma^2\left(\phi \;{\vec J}(\kappa) + (1-\phi){\vec I}\;\right)
\end{equation}

where $\sigma$ can be computed using:
\begin{equation}
	\sigma = \sqrt{\frac{{\vec r}^T {\vec C}^{-1}{\vec r}}{N}}
\end{equation}

Since $\sigma$ can be computed from the residuals, we only use our
slow numerical minisation algorithm need to find 
the value of $\phi$ \citep{Williams2008}. 

Note that
we only analysed 500 observations while nowadays time series with 7000
observations are not uncommon. If one tries to rerun our examples for this
value of $N$, then one will note this takes an extremely long time. The
main reason is that the inversion of matrix $\vec C$ requires  
\(\mathcal{O}(N^3)\) operations. \cite{Bosetal2008,Bosetal2013} have 
investigated how the covariance matrix $\vec C$ can be approximated by a 
Toeplitz matrix. This is a special type of matrix which has a constant value
on each diagonal and one can compute its inverse using only 
\(\mathcal{O}(N^2)\) operations. This method has been implemented in the 
Hector software package that is available from 
\verb!http://segal.ubi.pt/hector!.

The Hector software was used to create time series with a length of 
5000 daily observations (around 13.7 years)
for 20 GNSS stations which we will call the Benchmark Synthetic GNSS (BSG).
This was done for the 
the horizontal and vertical components, producing 60 time series in total.
Each contains a linear trend, an annual and a semi-annual signal. The sum
of flicker and white noise, $w_i$, was added to these trajectory models:
\begin{equation}
	w_i = \sigma\left[ \sqrt{\phi} \sum\limits_{j=0}^{i-1} h_{i-j} v_j +
					\sqrt{1-\phi} \;u_i\right]
\end{equation}

with both $u_i$ and $v_i$ are Gaussian noise variables with unit variance.
The factor $\phi$ was defined in Eq. (\ref{SimonRewrite}). 
To create our BSG time series we used $\sigma=1.4$ mm, $\phi=0.6$ and 
horizontal components and $\sigma=4.8$ mm, $\phi=0.7$ for the vertical
component.

It is customary to scale the power-law noise amplitudes 
by $\Delta T ^{-\kappa/4}$
where $\Delta T$ is the sampling period in years. For the vertical flicker 
noise amplitude we obtain:
\begin{equation}
 \sigma_{pl} = \frac{\sigma \sqrt{\phi}}{\Delta T ^{\kappa/4}} =
 		\frac{4.8 \times\sqrt{0.7}}{(1/365.25)^{1/4}} = 17.6 \;\;\text{mm/yr}^{0.25} 	
\end{equation}
 
The vertical white noise amplitude is 2.6 mm. For the horizontal
component these values are $\sigma_{pl}=4.7$ mm/yr$^{0.25}$ and 
$\sigma_w=0.9$ mm respectively.
The BGS time series can be found on the Springer website for this book, 
and can be used to verify the algorithms developed by the reader. These
series will also be compared with other methods in the following
chapters.

\section{Discussion}\label{discussion}

In this chapter we have given a brief introduction to the
principles of time series analysis.
We paid special attention to the maximum likelihood estimation (MLE) method 
and the modelling of power-law noise.  We showed that with our
assumptions on the stochastic noise properties, the estimated parameters have
their variance bounded by the Cramer Rao lower bound. Therefore the MLE is an
optimal estimator in the sense of asymptotically unbiased and efficient 
(minimum variance).

In this book
we will present other estimators such as the Kalman filter, the Markov Chain 
Monte Carlo Algorithm and the Sigma-method. All have their advantages and
disadvantages and to explain them was one of the reasons for writing this
book. The other reason was to highlight the importance of temporal correlated
noise. This phenomenon has been known for a long time but due to increased
computer power, it has now become possible to include it in the analysis of
geodetic time series. We illustrated how this can be done 
by various examples in Python 3 that highlighted some numerical aspects
that will help the reader to implement their own algorithms.

\end{document}